\documentclass[10pt,journal,compsoc]{IEEEtran}

\usepackage[T1]{fontenc}		
\usepackage[utf8]{inputenc}		
\usepackage{color}				
\usepackage{graphicx}			
\usepackage{microtype} 			
\usepackage{float}              
\usepackage{caption}

\usepackage{fancyhdr}
\usepackage{authblk}
\usepackage{xcolor,soul,framed} 
\colorlet{shadecolor}{yellow}
\usepackage{subfig,amsmath,cleveref,caption}
\ifCLASSOPTIONcompsoc
  \usepackage[nocompress]{cite}
\else
  \usepackage{cite}
\fi
\ifCLASSINFOpdf
 
\else
 
\fi

\begin{document}

\title{An Audio Envelope Generator Derived from Industrial Process Control}

\author{Ashwin Pillay}
\affil{University of Mumbai, Mumbai, India}
\affil{\textit {(2016.ashwin.pillay@ves.ac.in)}}

\markboth{Pillay}{An Audio Envelope Generator Derived from Industrial Process Control}

\IEEEtitleabstractindextext{%
\begin{abstract}
Audio envelopes serve a crucial role in ensuring the versatility of synthesizers in producing timbres. To this end, the Attack, Decay, Release and Sustain (ADSR) envelope generator and its derivatives have been established as a mainstay in modern music. However, there may be merit in exploring alternate techniques to produce envelopes that could not only resemble ADSR but also be used to create novel timbres. Consequently, an attempt is made in this research to formulate the framework of a new envelope generator by redefining the Proportional-Integral-Derivative (PID) algorithm used in feedback-based process control. Additionally, a detailed analysis is made on the modes of operation and the nature of envelopes thus generated to establish it as a potential harbinger of distinctive styles of music.
\end{abstract}}

\maketitle
\IEEEdisplaynontitleabstractindextext
\IEEEpeerreviewmaketitle

\section{INTRODUCTION}
\label{S0}
Ever since its advent, modern synthesizers have revolutionized the way music is produced. However, on its own, a mere triggering of the synthesizer keys may not faithfully recreate the cornucopia of sounds produced by physical instruments or those present in nature. As a result, the concept of audio envelopes was developed to control various aspects like the pitch, filter cutoff frequencies and most commonly, the amplitude of an electronically synthesized sound to increase the degrees of freedom involved with timbre generation. Envelope generators have therefore been an integral part of synthesizers right from their inception \cite{pinch}.

The technological advancements in digital electronics and computers ushered significant improvements in the portability, storage capabilities, and methods of sound generation available for synthesizers \cite{vail}. However, the ADSR envelope generator conceptualized initially as part of the Moog Synthesizer  \cite{pinch} in the 1960s continues to be a popular technique to create envelopes for both hardware and software synthesizers \cite{vail}. The main factors contributing to a wide acceptance are its relative simplicity in implementation and ability to model many musical instruments and natural sounds. While simplified variants like the attack-release (AR) and enhanced versions having the hold (AHDSR) or delay (DADSR) phases have been implemented for some synthesizers \cite{vail}, the fundamentals of ADSR envelope generation have remained unchanged to a great extent.

\begin{figure}
\centering
\includegraphics[width=0.5\textwidth]{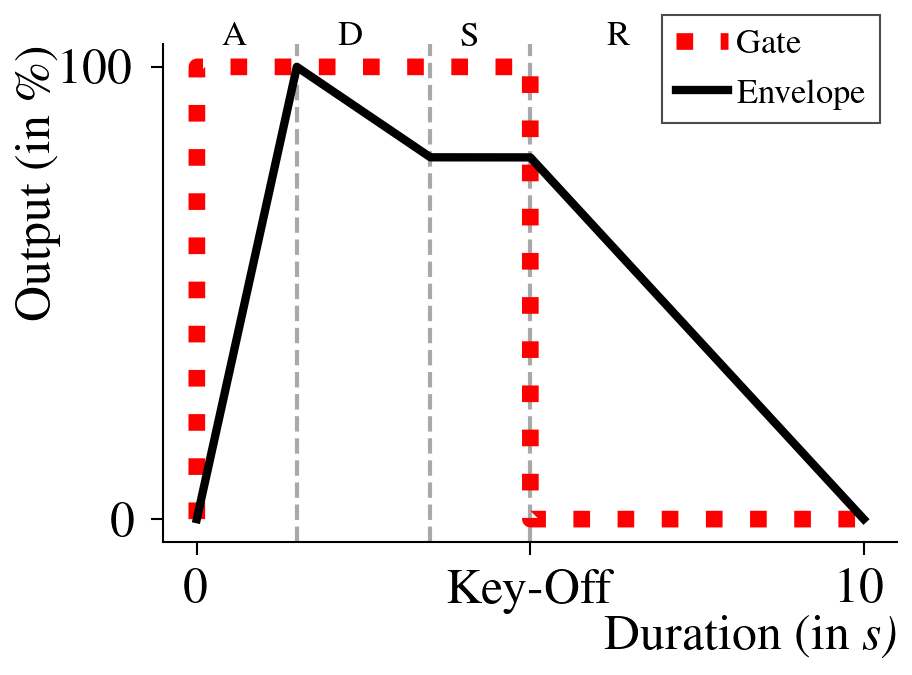}
\caption{The ADSR envelope generator has remained the most popular choice for synthesizer manufacturers owing to its relative simplicity and versatility. During key-on, the envelope sweeps through the attack, decay and sustain phase. On key-off, it undergoes the release phase.}
\label{F1}
\end{figure}

\subsection{PID Control Scheme}
\label{S01}

\begin{figure}[t]
\centering
\includegraphics[width=0.5\textwidth]{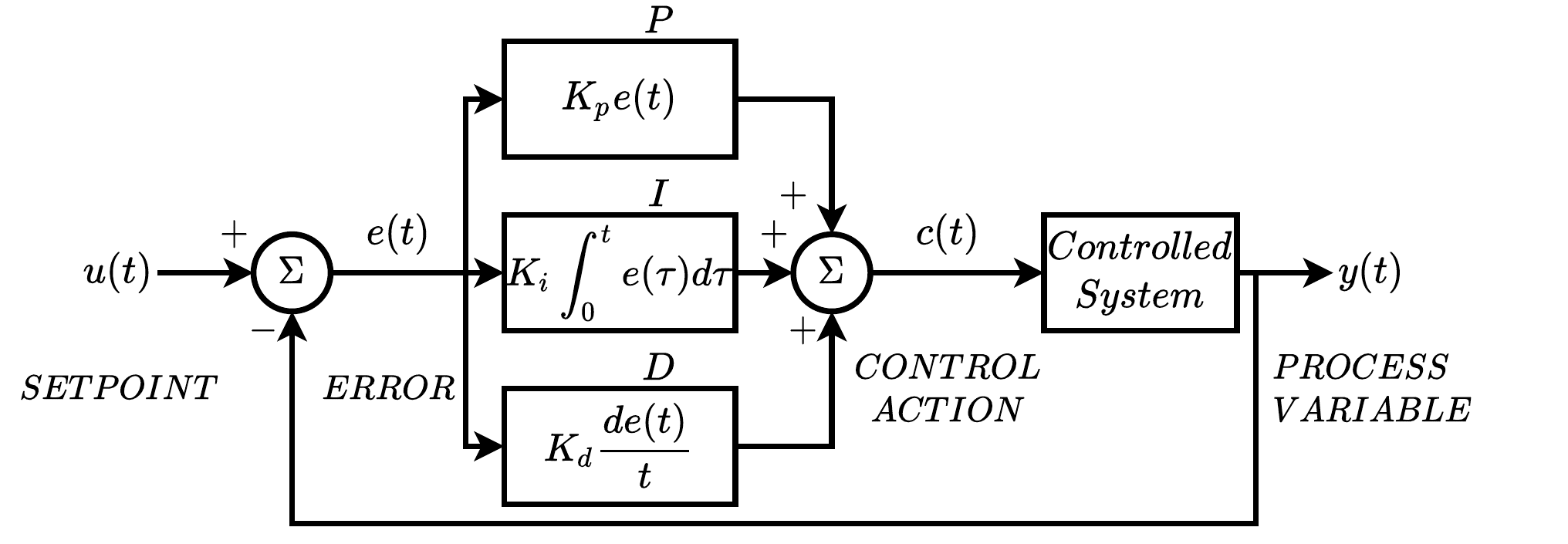}
\caption{PID control acts on the error between the desired setpoint and current output of the controlled system and is essentially a combination of three main components. While the P component operates only on the immediate error, the I and D components consider the past and future errors, respectively.}
\label{F2}
\end{figure}

The PID control strategy is considered among the most effective and widespread techniques used in industrial process control  \cite{lin}. It chiefly involves producing control action as feedback to the error (E) between the desired value or setpoint (SP) and the current value of the process variable (PV) to be controlled. The action produced effectively combines the control’s three main constituents: the proportional (P), integral (I) and derivative (D) components.

While the P component is dependent only on the instantaneous difference between SP and PV, the I component records historical trends by integrating such differences over time \cite{liptak}. Additionally, the D component anticipates the future course of the PV based on the rate of change of these differences. From a control theory perspective, the I component provides low-frequency compensation by adding a pole at zero, thereby regulating the steady-state error to nil \cite{ang}. Correspondingly, the D component adds high-frequency compensation by adding a pole at infinity, moderating the transient response of the controller. Ultimately, when all components are active, the PID control scheme is a form of lead-lag compensation.

Additionally, each component of PID is associated with a gain factor to adjust its contribution to the final control output; these are termed as $K_{p}$, $K_{i}$ and $K_{d}$, respectively. The efficacy of a PID control depends on the values chosen for this set of gains, and sometimes even on how these values change with time \cite{lee}. To this end, several tuning methods have been developed and are beyond the scope of this research. The overall transfer function for PID control may be obtained from \cref{F2} as \cref{E1} in the continuous-time domain and \cref{E4} in the discrete-time domain \cite{sreenivasappa}.

\begin{equation}
c(t) = K_{p}e(t) + K_{i}\int_{0}^{t}e(\tau).d\tau + K_{d}\frac{de(t)}{t}
\label{E1}
\end{equation}

\begin{equation}
\begin{aligned}
c[n] &=   c[n-1] + K_{p}\{e[n]-e[n-1]\} + \frac{K_{i}}{f_{s}}e[n] \\
&+ K_{d}f_{s}\{e[n]-2e[n-1]+e[n-2]\} \\
&where: f_{s} = sampling\ frequency
\end{aligned}
\label{E4}
\end{equation}

While the versatility of the PID scheme has led to widespread industrial acceptance, its applications are mostly confined to being a means of achieving continuous control. There is a dearth of significant research on non-control related PID implementations in academia, posing whether such possibilities are realistic or surprisingly overlooked.

Similarly, the availability and versatility of synthesizers heralded new genres of music— often going beyond the possibilities of what classical musical instruments could produce \cite{collins}. While this led to the growing use of ADSR envelope generators in experimental endeavours, the fact that the core of the ADSR technique has withstood notable innovation might have impeded an escalation to more intriguing forms of audio synthesis. Consequently, an effort is made in this study to analyze the potentialities of a novel form of audio envelope generation, fusing two technologies whose fundamentals have withstood the wave of innovation around them: a PID based audio envelope generator, hereafter referred to as PIDEG.

This paper is organized as follows. \Cref{S1} establishes the link between PID control and envelope generation and subsequently lays out the PIDEG framework. \Cref{S2} discusses some of the possible design choices to be made while implementing PIDEGs in practice. \Cref{S3} analyzes the working of PIDEGs in detail, including the possible modes of operation and the envelope shapes generated for each of them. \Cref{S4} summarizes the study after presenting some of the potential research directions in the future. Finally, \cref{S5} culminates the study.

\section{The PID Envelope Generator}
\label{S1}

\begin{figure}[!b]
\centering
\includegraphics[width=0.5\textwidth]{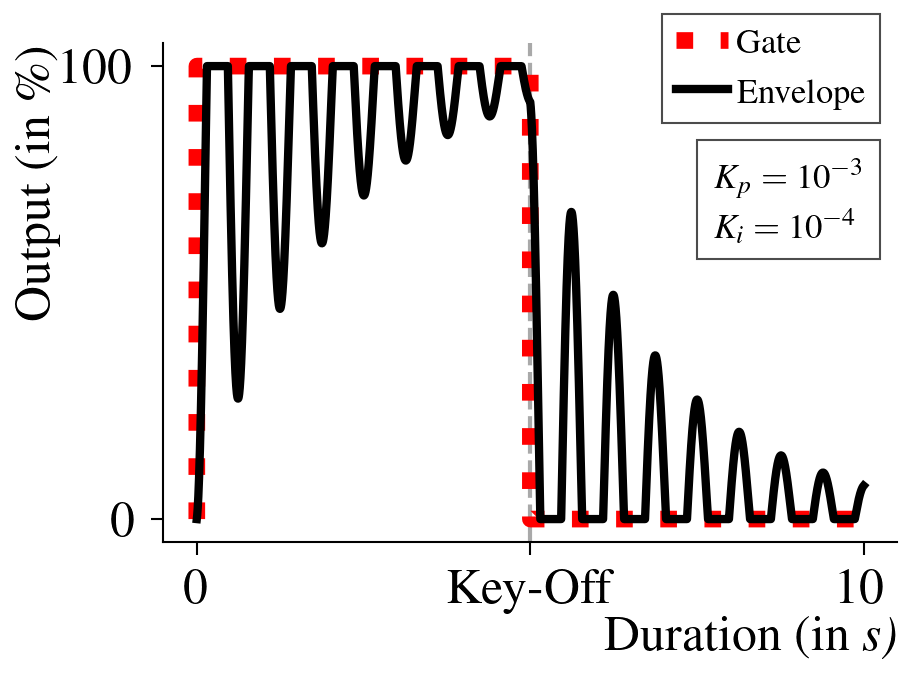}
\caption{PIDEGs extend the concept of PID control for generating audio envelopes. Like the ADSR, the envelope generation is controlled by the gate signal representing key-on when active and key-off otherwise. The shape of the envelope thus generated can be controlled by its five parameters: $K_{p}$, $K_{i}$, $K_{d}$, $K_{r}$ and $K_{f}$.}
\label{F3}
\end{figure}

The premise of PID based envelope generation lies in redefining the control scheme as a mechanism to bring the value of one entity closer to another— traditional PID controllers have a fixed SP value against which it generates an output in an attempt to get the PV value increasingly closer to SP. Audio envelopes, on the other hand, are governed by gate signals: a binary square wave corresponding to the synthesizer “key-on” when it is triggered and “key-off” otherwise. 

In almost all cases, during key-on, the envelope tries to bring the aspect of sound under control (usually amplitude) towards the maximum value (100 \%). Correspondingly, it brings this aspect back towards the minimum (0 \%) when the key is off, as evidenced by \cref{F1}. In PID terminology, this may be interpreted as setting SP to 100 \% on key-on and 0 \% on key-off, the resulting controller output being the envelope generated. Thus, by setting desired values of $K_{p}$, $K_{i}$ and $K_{d}$ parameters, the shape and duration of this envelope can be controlled in a way analogous to that provided by ADSR parameters: this lays the crux of PIDEG, a PID controller reformed to generate audio envelopes (see \cref{F3}).

Deviating from the orthodox use of PID for feedback control, PIDEGs are robust enough to account for variations in SP value over time— the output produced by the controller at any instant of time is calculated for the instantaneous SP. This allows the use of continuous or piecewise setpoint curves to generate envelopes with increased granularity. Hence, in the generalized definition of PIDEG, the SP is referred to as the “leader” curve LC, while the envelope output is termed as the “follower” curve FC. The integration of LC adds two more control parameters to the envelope generator: the rise gain ($K_{r}$) and fall gain ($K_{f}$) factors that regulate the time taken by the LC to reach 100 \% on key-on and 0 \% on key-off, respectively.

\section{Implementational Choices}
\label{S2}

The core concept of PIDEG defined previously is in a generalized form, leaving many degrees of freedom for audio engineers to develop varying kinds of PIDEG implementations. In this section, some of the potential areas where such design choices can be made are discussed.

\subsection{LC Functions}
\label{S21}

\begin{figure}[!b]
\centering
\includegraphics[width=0.5\textwidth]{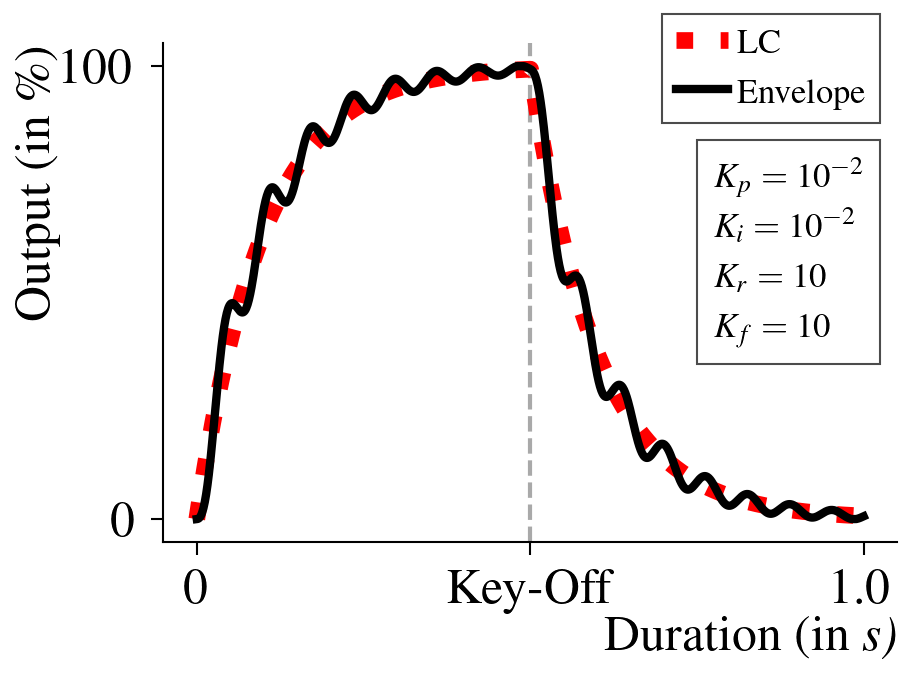}
\caption{The inverse exponential function results in zero at infinity. Hence, it can be used without safeguards to maintain SPs within range, allowing simplicity in design.}
\label{F5}
\end{figure}

Theoretically, there are no limitations to the kind of LCs against which a PIDEG can generate envelopes, providing considerable encouragement for experimentation in synthesizing timbres that may be beyond the possibilities of the conventional ADSR. However, it must be noted that the envelope output should be restricted within the range of 0-100 \%. For most implementations, an LC having its value in this range at all times would be sufficient to allow for simplicity in design without compromising on the variety of envelope shapes that could be generated using it. Hence, the inverse-exponential function, given by \cref{E5}, is suggested in this study as the stock LC generating function for the PIDEG—ensuring that a sustained key-on will cause the output to saturate at 100 \% while on key-off, it will eventually drop to 0 \%. The LC generated using this function can be visualized by \cref{F5}.

\begin{equation}
LC(n)  = \left\{ \begin{array}{l}
  1-e^{-\frac{K_{r} \times n}{f_{s}}},\ \textrm{during } key-on \\[4pt]
 e^{-\frac{K_{f} \times n}{f_{s}}},\ otherwise\\
 \end{array} \right.
\label{E5}
\end{equation}

\begin{figure}[t]
\centering
\includegraphics[width=0.5\textwidth]{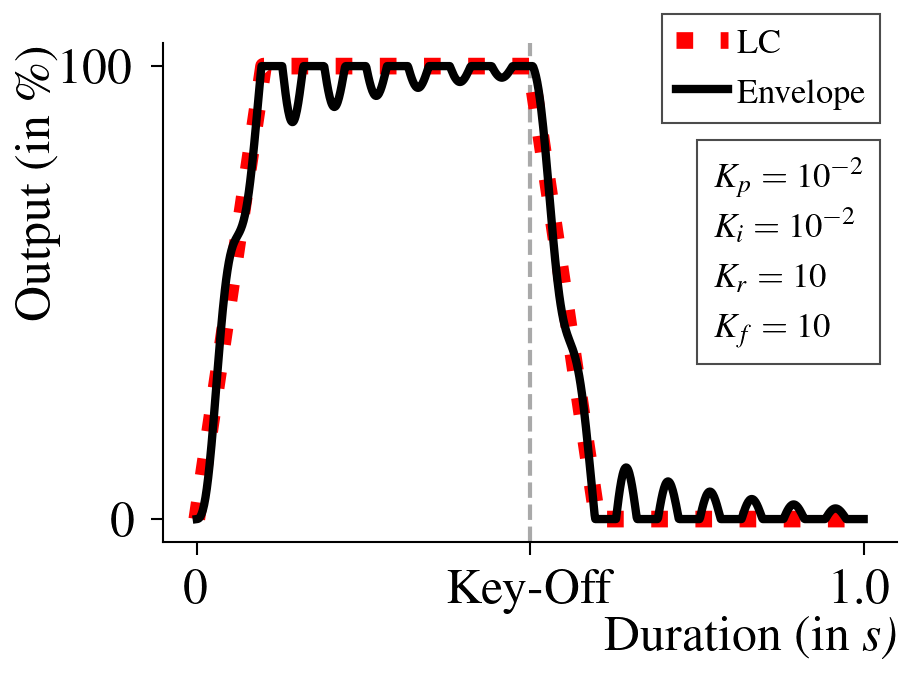}
\caption{The trapezoidal function is suited for generating LCs in systems without floating-point units. However, its linear rise and fall tends to limit the variety of envelope shapes produced.}
\label{F6}
\end{figure}

The trapezoidal function represented by \cref{E6} is another suitable candidate for LC generation. Compared to inverse-exponential LCs, they might not lead to a wide variety in the PIDEG envelopes produced. However, being devoid of floating-point computations, generating trapezoidal LCs may be more feasible for low-resourced microcontrollers. The envelope generation, in this case, is depicted in \cref{F6}. Additionally, there exist numerous other LC generation functions, including the boxcar \cite{weisstein_boxcar} and staircase functions \cite{weisstein_staircase}, that are beyond the scope of this study.

\begin{figure}[!b]
\centering
\includegraphics[width=0.5\textwidth]{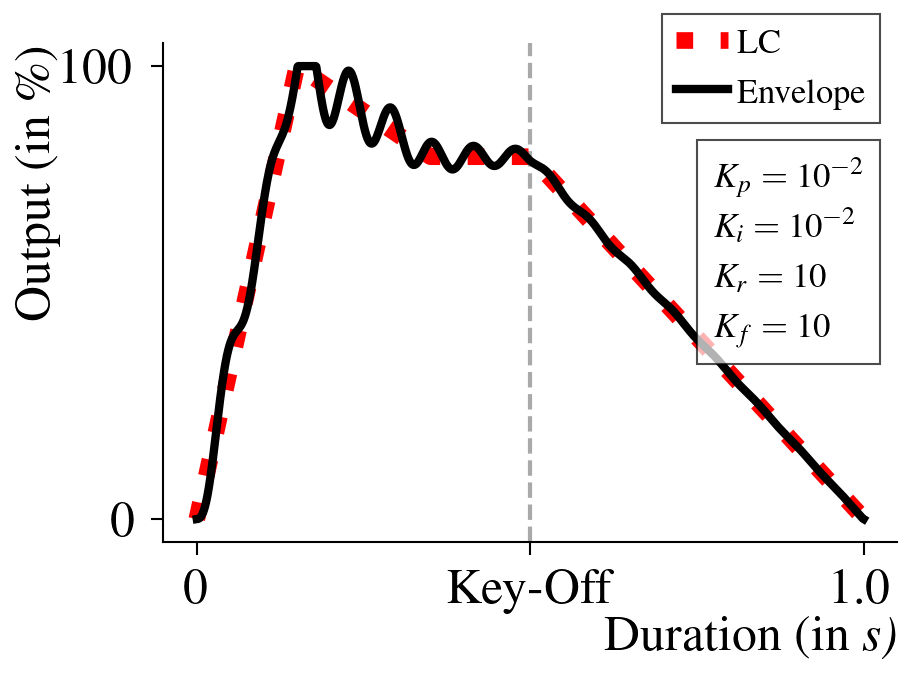}
\caption{By patching it to the PIDEG, the ADSR envelope generator can be used as an LC source. The PID induced oscillations present in the envelope thus generated can be used to better model tremolo effects. Whenever LC is supplied externally, the $K_{r}$ and $K_{f}$ PIDEG parameters are inactive.}
\label{F8}
\end{figure}

\begin{equation}
LC(n)  = \left\{ \begin{array}{l}
 \min(1, \frac{K_{r} \times n}{f_{s}}),\ \textrm{during } key-on \\[4pt]
 \max(0, 1 - \frac{K_{f} \times n}{f_{s}}),\ otherwise\\
 \end{array} \right.
\label{E6}
\end{equation}

For all aforementioned cases, the PIDEG parameter $K_{r}$ is used to control the rate with which the LC “rises” towards 100 \% on key-on. Correspondingly, the $K_{f}$ parameter is used during key-off to control how quickly it “falls” back to 0 \%— higher the value of these parameters, faster is the respective rise and fall.

Additionally, PIDEGs can also allow externally provided signals to become their LC source. The PID envelope generated by using an ADSR envelope as its LC is illustrated in \cref{F8}. Multiple envelope generators can be thus chained in series to obtain many varied and unprecedented envelope shapes. When external signals are used as LCs, the $K_{r}$ and $K_{f}$ parameters of PIDEG are deactivated and their set values are disregarded.

\subsection{Integral Windup}
\label{F22}

\begin{figure}[!b]
\centering
\subfloat[Envelope generated for PIDEG having integral windup]{\includegraphics[width=0.25\textwidth]{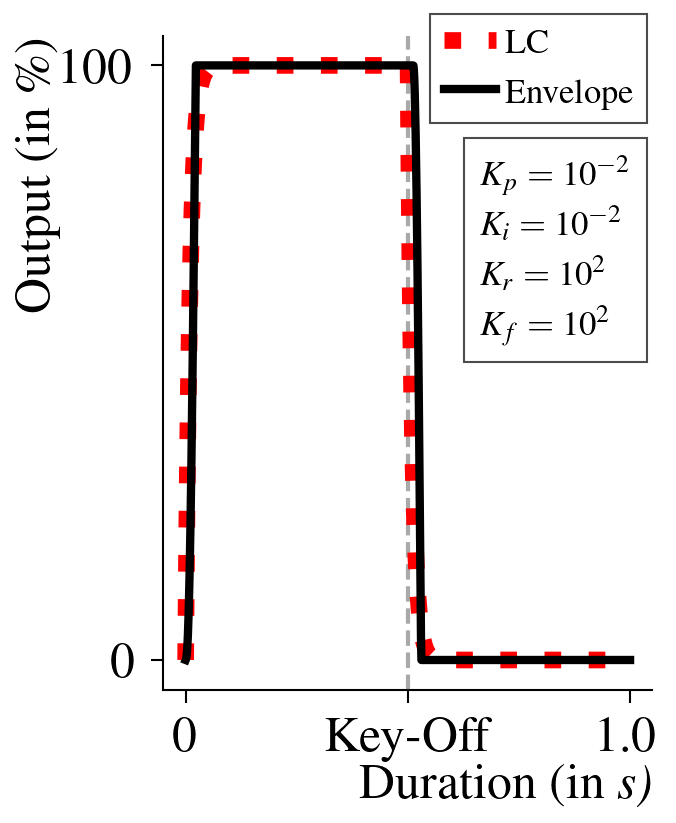}\label{F9a}}
\subfloat[Envelope generated without integral windup for the same set of parameters]{\includegraphics[width=0.25\textwidth]{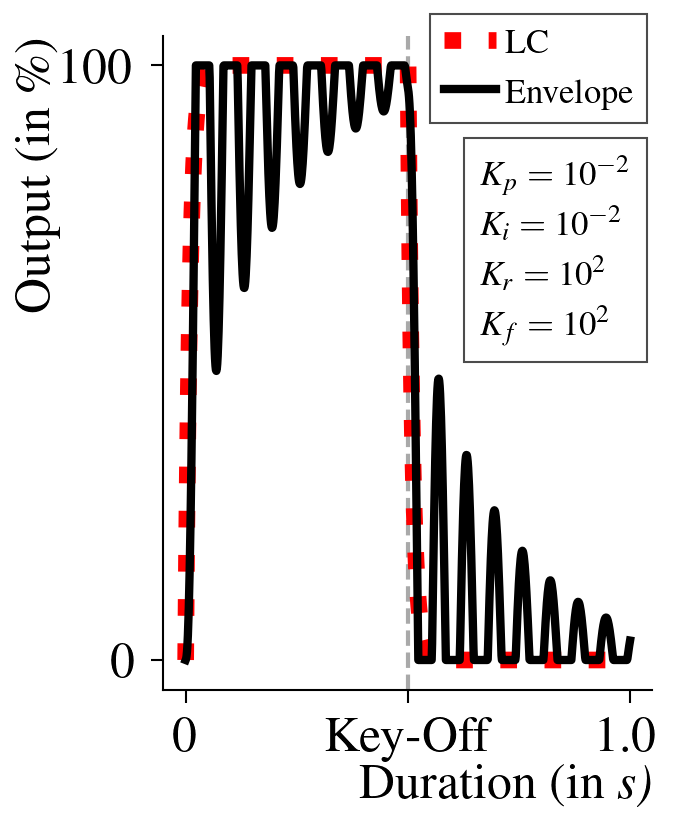}\label{F9b}}
\caption{Integral windup has the effect of steadily maintaining the envelope at either limit. For the same set of parameters, the envelopes generated by PIDEGs operating without windup contain oscillations with decreasing amplitudes about these limits. 
}
\label{F9}
\end{figure}

Like its counterpart in the process control domain, PIDEGs are also liable to integral windup  \cite{shin}— the envelope output is constrained to be strictly between 0-100 \%, both inclusive. While such range limitations are problematic in control applications, they may be desirable in the case of PIDEGs. Integral windups will cause the envelope output to be constantly clamped to 100 \% during prolonged durations of key-on and to 0 \% on prolonged key-off as described by \cref{F9a}. The absence of integral windup, on the other hand, will cause oscillations in the PIDEG output about 100 \% and 0 \% on sustained key-on and key-off, respectively, as evidenced by \cref{F9b}.

Therefore, incorporating integral windup is the preference of implementers and musicians, depending upon the properties they desire in the envelope output. For instance, using PIDEGs having windup might be better suited for the generation of ambient drones. Contrarily, a PIDEG free of this property may be more amicable to model tremolo effects of string instruments like the violin. By nature, PIDEGs possess integral windup by default. However, a convenient way to avoid it is by isolating the FC output from that of the envelope. In such situations, the FC is free to assume any value resulting from per-sample PID calculations. At the same time, a hard limiter will ensure this value is within the valid range before appearing as the envelope output (EO) as described by \cref{E7}.  While this technique seems to allow unrestricted accumulation of the integral in theory, the range of computer data units (in software) and capacitor charge limits (in analogue electronics) tend to, at best, defer the effects of windup in practice. As an alternative, any of the widely accepted anti-windup solutions \cite{da_silva} can also be implemented. Furthermore, software implementations of PIDEGs may include a switch to allow users to activate integral windup as per their requirement.

\begin{equation}
EO(n) = \max(0, \min(1,FC(n))
\label{E7}
\end{equation}

\subsection{Bumps in Envelope Output}
\label{S23}

“Bumps” in controller output \cite{cheong} is another undesirable phenomenon in PID control that is also observable for PIDEGs. They are likely to occur when there are quick and significant changes in the LC values, like during integral windup or for step-function LCs. However, addressing these bumps is also subject to the implementor’s liking, and most of the commonly available bumpless transfer techniques should prove remedial \cite{cheong}. PIDEGs that only support continuous function LCs include bumpless transfer through this constraint itself, and consequently, the bumps in the envelopes generated are negligible.

\subsection{Negotiating Key-Off Instability}
\label{S24}

\begin{figure}[t]
\centering
\subfloat[Runaway envelope generated as a result of instability in PID control. Note that the oscillations in key-off are sustained and hence, never-ending.]{\includegraphics[width=0.25\textwidth]{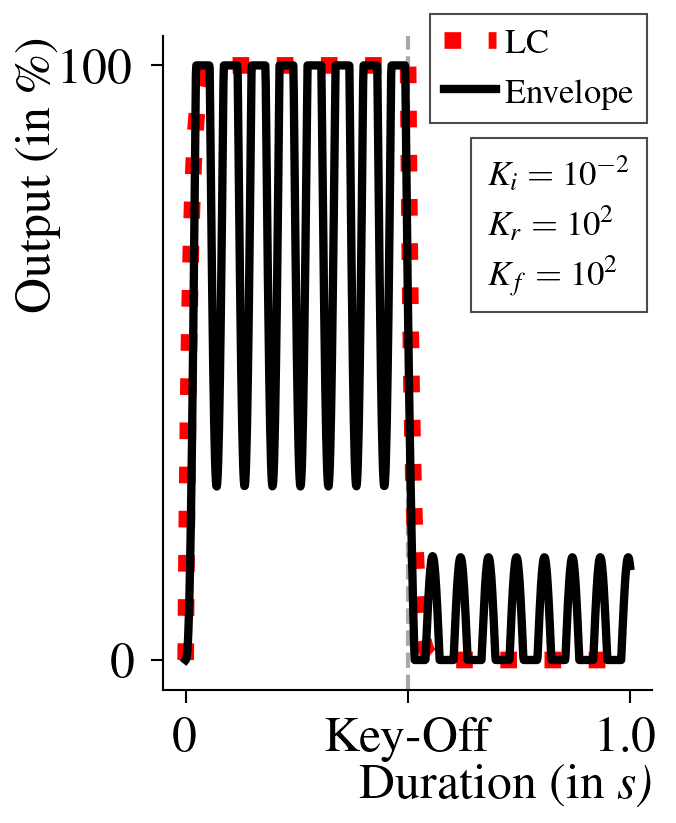}\label{F4a}}
\subfloat[The envelope when generated with takeover mechanism kicking-in at 0.6 s]{\includegraphics[width=0.25\textwidth]{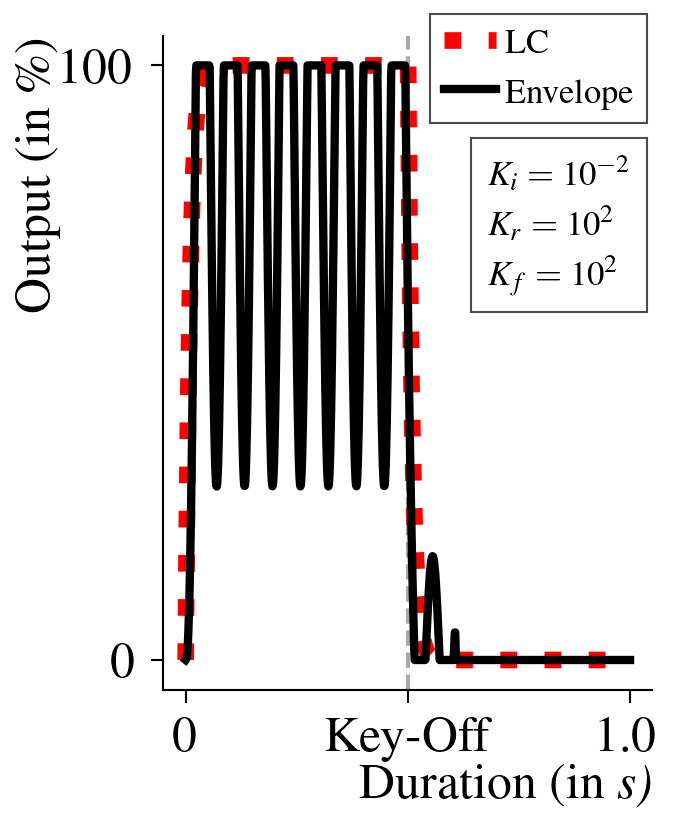}\label{F4b}}
\caption{Similar to its applications in industrial control, the PID algorithm is prone to instabilities. In the case of PIDEGs, these appear as infinite-length envelopes during key-off. Implementing a takeover mechanism can prevent this by ensuring that envelopes eventually decay back to 0 \%.}
\label{F4}
\end{figure}

Similar to the observations in feedback control \cite{liptak}, the P component in PIDEGs tends to introduce instability conditions for gains higher than the ultimate gain. The D component, being sensitive to sudden changes, is also likely to do the same at higher $K_{d}$ values. Meanwhile, the nature of the I component will always lead to unstable output, as discussed in \cref{S31}.  Though instability may not be of much concern during the key-on phase and might instead be a preference, it could lead to a potentially never-ending envelope being generated during the key-off phase, as demonstrated in \cref{F4a}. Hence, efforts must be made to ensure that PIDEGs always result in envelopes that will drop to 0 \% without compromising the effects of the control parameters during key-off.

Some of the possible ways of reducing the chances of instability are limiting the maximum gain set for $K_{p}$, $K_{i}$ and $K_{d}$ to one and turning off the I and D modes after some time into key-off. However, as observed in \cref{F12}, it might not be enough to prevent this situation. Another option would be implementing a “takeover mechanism”, a preset array of PIDEG outputs generated once a specific condition has been met during the key-off phase to ensure that the envelope eventually decays to 0 \%. Particulars like the total number and values of the takeover samples and the condition at which they must be triggered are left to the implementer’s choice. For example, outputting an inverse-exponential stream of 250 envelope samples once the total number of samples generated in the key-off phase has exceeded $K_{f}$, and the envelope output has fallen under 2 \%, as illustrated in \cref{F4b}.

\section{Analysis of Envelopes Generated}
\label{S3}

To demonstrate the effect of each control parameter on the envelope shape and the modes of operations resulting from their permutations, a computer-program version of PIDEG was implemented in C language. Envelopes were generated at a sampling rate of 1 kHz, against inverse-exponential LCs having $K_{r}$ = 100 and $K_{f}$ = 100 (as seen in \cref{F11}), the integral windup being deactivated for each PID mode examined. Additionally, the D component was also turned off after 92 samples into key-off to reduce the probability of instability.

There may be endless possibilities for generating envelopes in each mode. However, only a subset is considered to understand PIDEG working in each of them as part of this section.

\subsection{Single-controlled Modes}
\label{S31}

\begin{figure}[t]
\centering
\subfloat[$K_{p} = 10^-4$]{\includegraphics[width=0.25\textwidth]{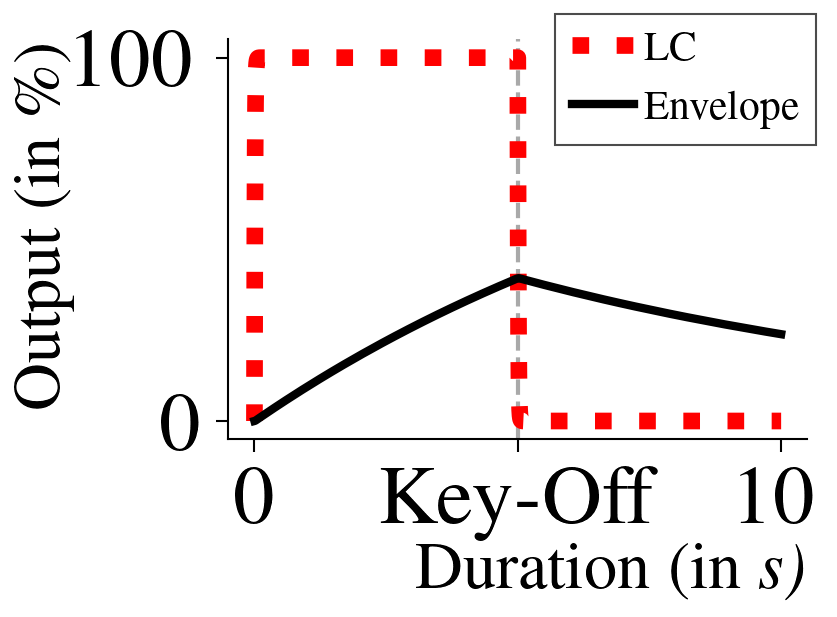}\label{F11a}}
\subfloat[$K_{p} = 10^-3$]{\includegraphics[width=0.25\textwidth]{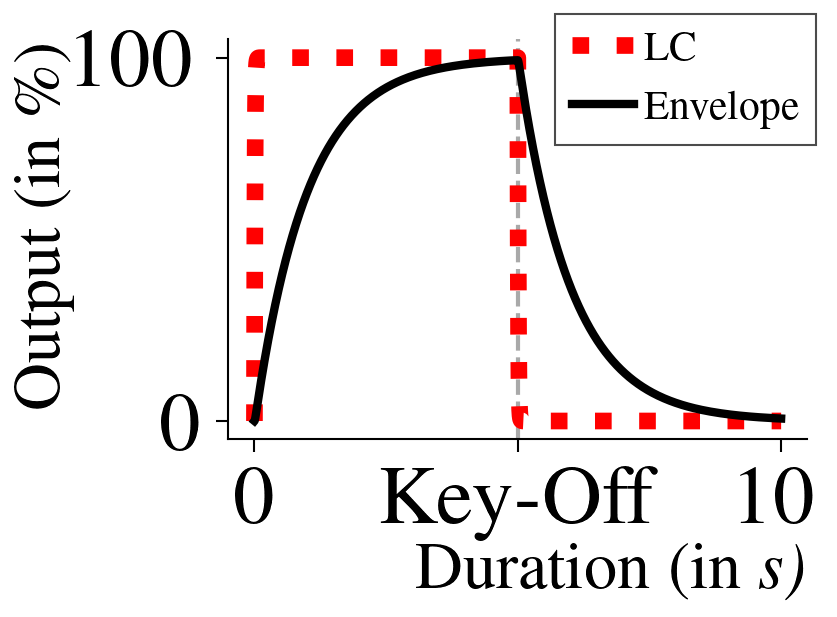}\label{F11b}}\hfill
\subfloat[$K_{p} = 10^-1$]{\includegraphics[width=0.25\textwidth]{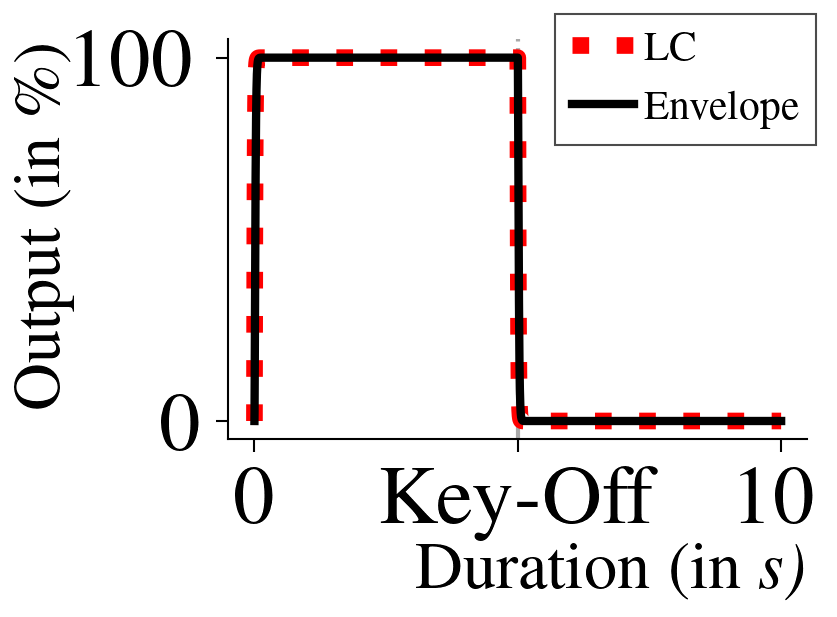}\label{F11c}}
\subfloat[$K_{p} = 2$]{\includegraphics[width=0.25\textwidth]{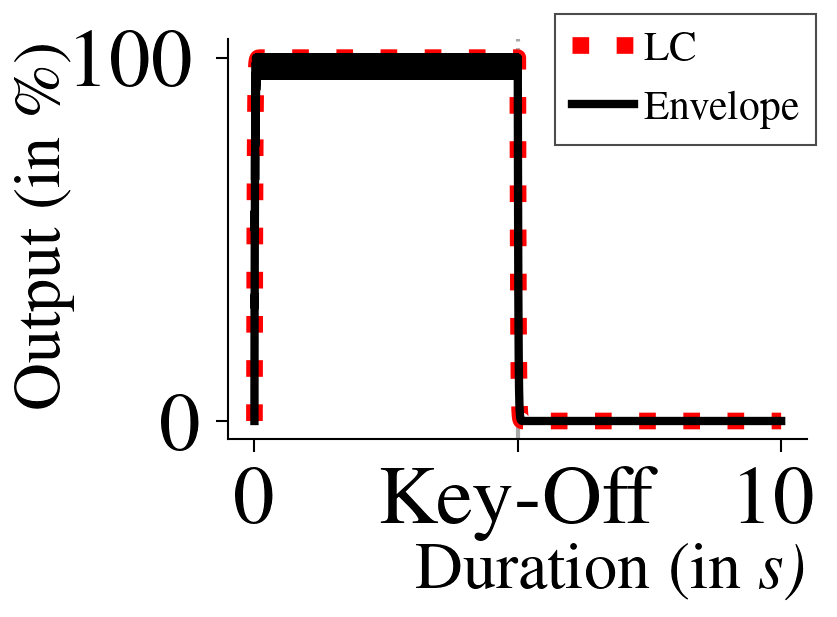}\label{F11d}}\hfill
\caption{Envelopes generated for PIDEGs operating in P-mode, against LC having $K_{r} = K_{f} = 100$. As the value of $K_{p}$ is increased, the FC follows LC more aggressively. At values beyond one, the envelope consists of  high-frequency oscillations denoted by the black bars in \cref{F11d}}
\label{F11}
\end{figure}

At an elementary level, PIDEG can operate with only one of its PID controls activated at a time. In the P-mode, the $K_{i}$ and $K_{d}$ parameters will be set to zero, and $K_{p}$ will be regulated to understand its effect on the envelope shape. \Cref{F11} represents the variations in envelope shape as this parameter is varied from $10^{-4}$ to 2. 

Since P-mode is directly proportional to the error between SP and PV, the value of $K_{p}$ effectively determines the speed of overall rise and fall of the envelope on key-on and key-off, respectively. Compared with the ADSR envelope, the values of $K_{p}$ and $K_{r}$ as a whole are analogous to the attack time, while $K_{p}$ and $K_{f}$ together correspond to the release time. As observed in \cref{F11a}, a lower value of this parameter leads to a higher attack and release time. Contrarily, on setting higher values, the FC will tend closer to the value of  LC and will have the same output eventually, as evidenced in \cref{F11c}. Any further increase in $K_{p}$ will cause FC to oscillate about LC at half the sampling frequency as discussed in \cref{S24} and seen for \cref{F11d}. While this may lead to the generation of desirable timbres, implementers must ensure that instability conditions are avoided by keeping a ceiling on $K_{p}$ values and preventing infinite length envelopes from being produced as a result.

\begin{figure}
\centering
\subfloat[$K_{i} = 10^-4$]{\includegraphics[width=0.25\textwidth]{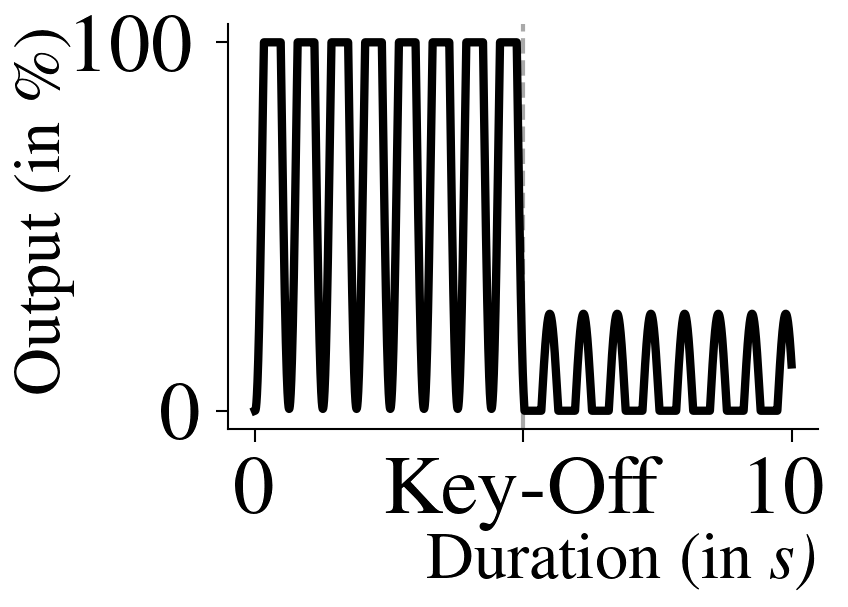}\label{F12a}}
\subfloat[$K_{i} = 10^-3$]{\includegraphics[width=0.25\textwidth]{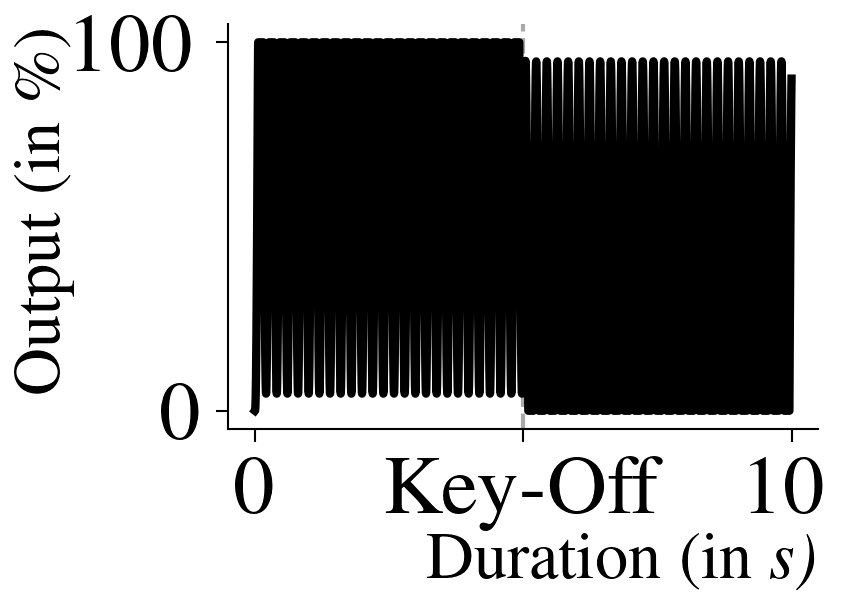}\label{F12b}}\hfill
\subfloat[$K_{i} = 10^-2$]{\includegraphics[width=0.25\textwidth]{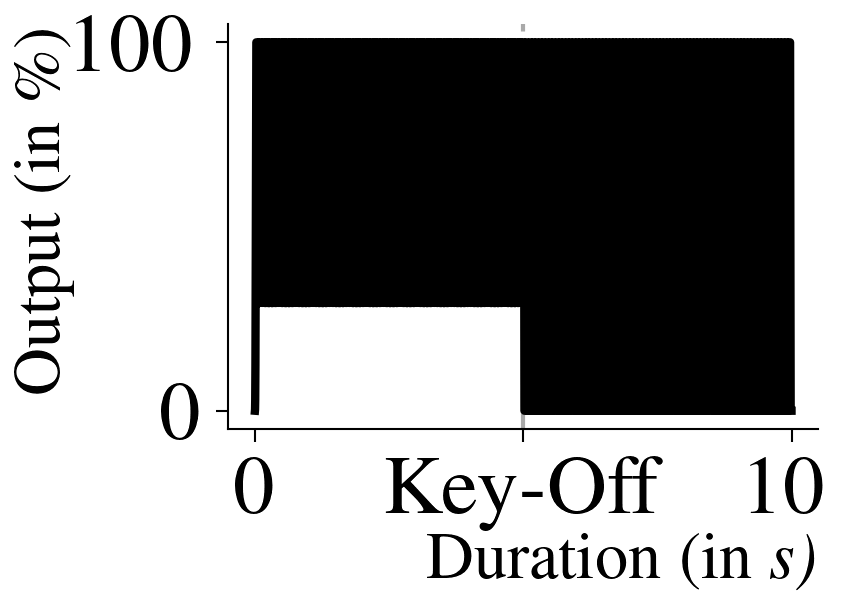}\label{F12c}}
\subfloat[$K_{i} = 10^-1$]{\includegraphics[width=0.25\textwidth]{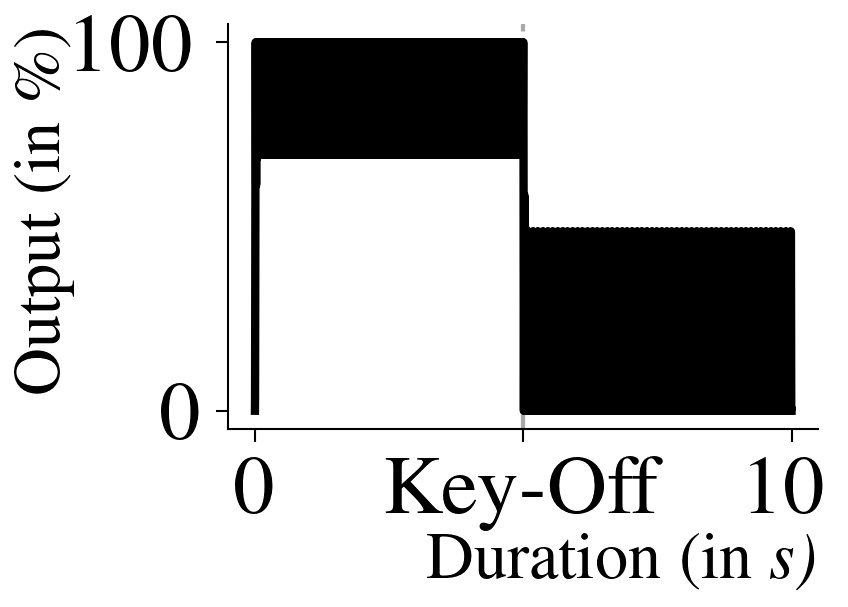}\label{F12d}}\hfill
\caption{Envelopes generated for PIDEGs operating in I-mode. As the value of $K_{i}$ is increased, the oscillations in the envelope produced tend to be of higher frequency (represented by the black bars) and decreasing amplitudes. Note that residue error integrals carried over from key-on may make this observation less apparent during key-off.}
\label{F12}
\end{figure}

In the I-mode, $K_{i}$ will be the sole active PIDEG parameter in operation. A phase lag is introduced to the FC thus generated and is hence sluggish in following the LC. It overcorrects the errors accumulated while lagging behind it and consequently collects an equivalent amount of error while leading. This results in oscillations occurring in the envelope about the instantaneous value of LC, with the overall output resembling FM waves as a consequence. The oscillatory nature of envelopes in this mode will lead to infinite-length envelopes, and hence, it should be used judiciously.

At lower values of $K_{i}$, the oscillations are higher in amplitude (in terms of displacement from LC) and lower in frequency; when the parameter is tuned to higher values, their amplitude decreases while frequency increases, as represented by the black bars seen in \cref{F12}. The explanation for this observation is that at higher values, the effect of integrating errors over time decreases, with the integral error dropping increasingly closer to zero. Therefore, the output is chiefly dependent on the instantaneous difference between LC and FC, meaning that the controller is effectively acting in the P-mode. In these situations, the $K_{i}$ value is comparable to $K_{p}$. Although, it must be noted that during key-off, it may be difficult to infer this observation since the integral and starting point of envelope samples depends on the last sample of key-on. It can significantly influence the ensuing envelope shape, as is the case for \cref{F12b} and \cref{F12c}.

\begin{figure}[t]
\centering
\includegraphics[width=0.5\textwidth]{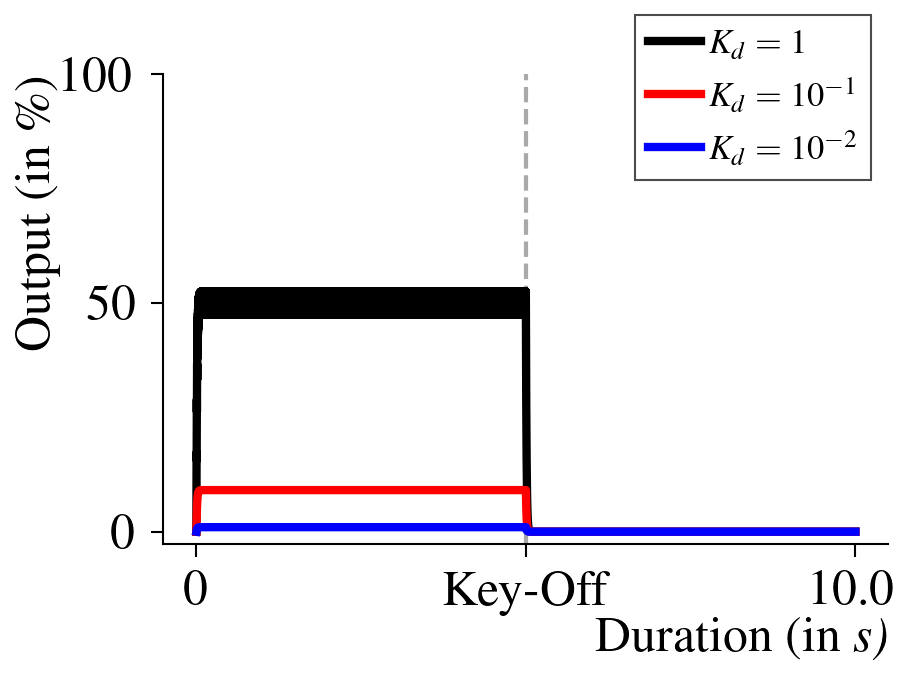}
\caption{Envelopes generated for PIDEGs operating in D-mode.  At lower values of $K_{d}$, the envelopes produced are weaker. However, on increasing its value beyond one, the resulting envelope contains oscillations about 50 \% at half the sampling frequency, as indicated by the black bars for $K_{d} = 1$.}
\label{F13}
\end{figure}

Finally, PIDEG operates in D-mode when only the $K_{d}$ parameter is activated. Here, the envelope shape largely follows the LC curve and is a miniature version of it at lower $K_{d}$ values like in \cref{F13}. This is because as the LC value saturates to 100 \%, the differences in consecutive errors (or the first-order derivative of the error) decay to 0, saturating the envelope output in the process, the value of saturation being directly proportional to $K_{d}$. 

However, at higher $K_{d}$ values, the control action produced is large enough to prevent this saturation of the envelope output. In fact, the envelope consists of triangle-wave oscillations occurring at half the sampling rate as it alternates between two values that average at 50 \%. However, since this implementation of PIDEG involves turning off D-mode after some time into key-off, such oscillations are not observed as the envelope decays back to zero.

\subsection{Dual-controlled Modes}
\label{S32}

\begin{figure}[!b]
\centering
\subfloat[]{\includegraphics[width=0.25\textwidth]{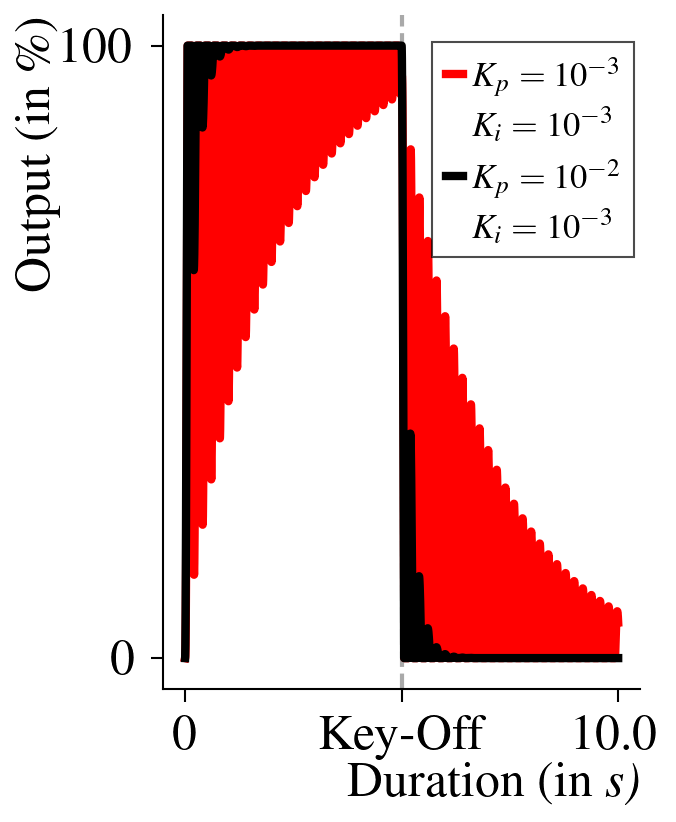}\label{F14a}}
\subfloat[]{\includegraphics[width=0.25\textwidth]{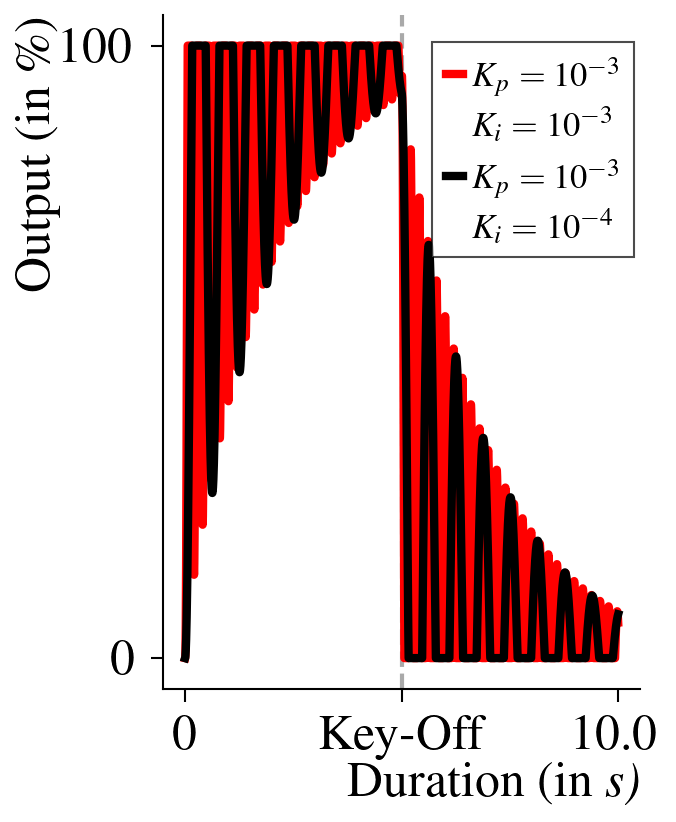}\label{F14b}}
\caption{Envelopes generated for PIDEGs operating in PI-mode. In this case, $K_{p}$ controls how fast the envelope follows the LC. Meanwhile, $K_{i}$ controls the frequency and amplitude of the oscillations present in the envelope similar to the I-mode.}
\label{F14}
\end{figure}

In dual-controlled modes, PIDEGs operate with any two of the PID parameters activated.  The resulting envelopes tend to amalgamate the effects of the individual controls that constitute each mode, with their specific contribution being proportional to the applied values.

When only $K_{p}$ and $K_{i}$ are varied, the PIDEG is in PI-mode. The P component, in this case, shapes the skeleton of the envelope: at higher values of $K_{p}$, it grows faster towards the LC, as observed in \cref{F14a}. Meanwhile, the I component controls the frequency of the oscillations in the envelope about the skeleton. Similar to the I-mode, these oscillations are of higher frequency as $K_{i}$ increases, like in \cref{F14b}. However, one crucial difference here is that the P component negates the phase lag inertia induced by the I component. Hence, the impact of  $K_{i}$ on the amplitude of oscillations is negligible. The PI-mode for PIDEGs, much like its counterpart in PID control, guarantees that the controlled value will eventually match the SP \cite{liptak}. Hence, it is desirable to generate audio envelopes in this mode to ensure that they settle to 100 \% and 0 \% without undergoing perpetual oscillations about these limits.

\begin{figure}[t]
\centering
\subfloat[]{\includegraphics[width=0.25\textwidth]{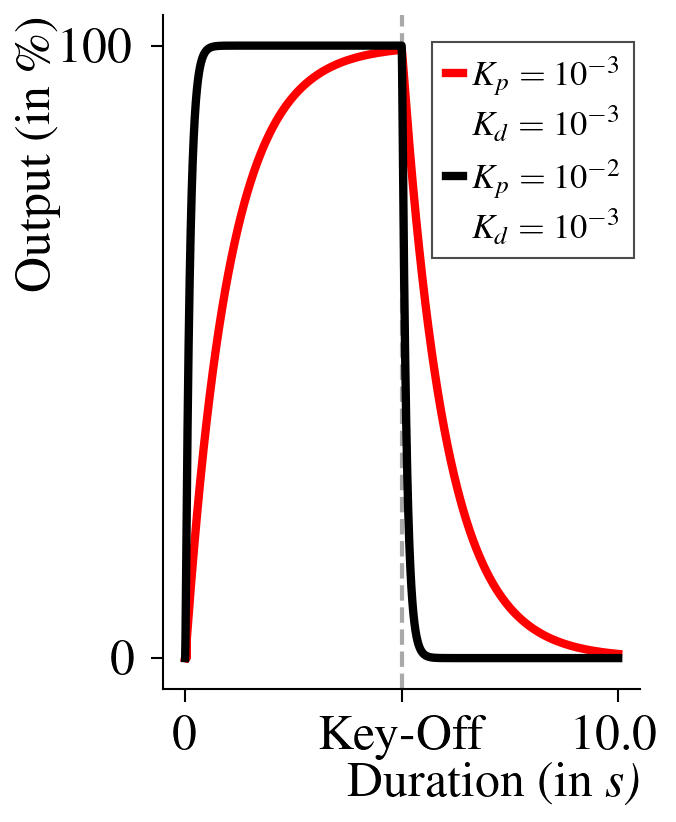}\label{F15a}}
\subfloat[]{\includegraphics[width=0.25\textwidth]{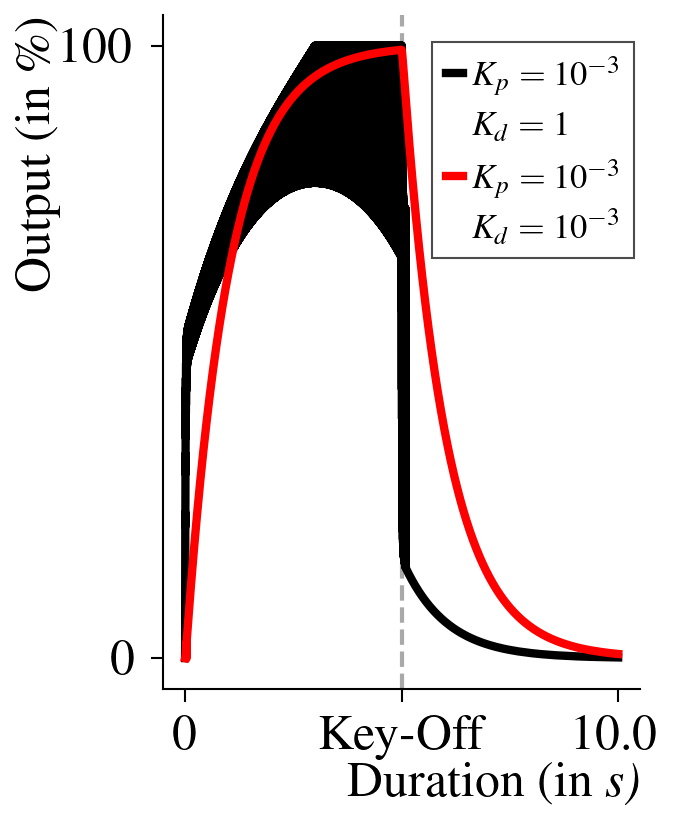}\label{F15b}}
\caption{Envelopes generated for PIDEGs operating in PD-mode. When $K_{d}$ is low, the effect of the D component is negligible, and the generator largely works in the P-mode. However, as $K_{d}$ is set to values beyond one, high-frequency oscillations are induced in the envelope. Additionally, there are sudden spikes that appear at the points of transition between key-on and key-off.}
\label{F15}
\end{figure}

In the PD-mode, where only $K_{p}$ and $K_{d}$ are active, the effect of the P component is similar to that of the PI-mode in shaping the envelope structure. However, the influence of the D component on the output is negligible at lower values of $K_{d}$, as is the case for \cref{F15a}; in these situations, the PIDEG operation is equivalent to P-mode. As $K_{d}$ is increased to values greater than one, the D component kicks in sharply, producing conspicuous bumps on significant changes in LC values (e.g., while transitioning from key-on to key-off) and high-frequency oscillations in the envelope generated.

\begin{figure}[h]
\centering
\subfloat[]{\includegraphics[width=0.25\textwidth]{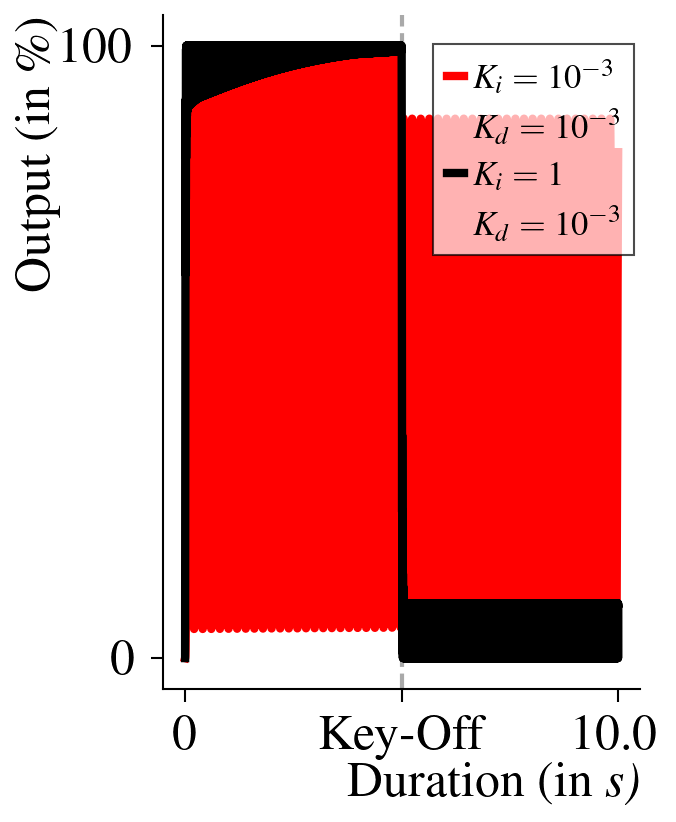}\label{F16a}}
\subfloat[]{\includegraphics[width=0.25\textwidth]{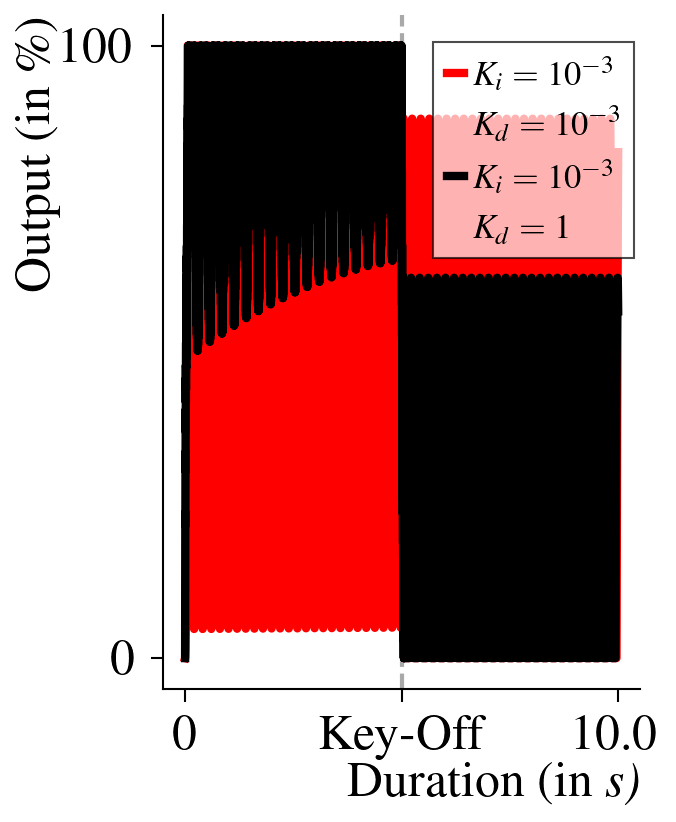}\label{F16b}}
\caption{Envelopes generated for PIDEGs operating in ID-mode. Here, the $K_{i}$ and $K_{d}$ parameters function similar to their working in the PI- and PD-modes, respectively.}
\label{F16}
\end{figure}

The ID-mode combines the effects of I and D components observed in the PI- and PD-modes, respectively. At lower $K_{d}$ values, this mode is similar to the I-mode as observed in \cref{F16a}. On increasing $K_{d}$, the D component begins to growingly oppose the lag introduced by the I component. As a result, the envelope output changes steeply whenever there occurs a sudden shift in LC like in \cref{F16b}, causing it to oscillate at the limits faster than it would for the equivalent I-mode. Unlike the PI-mode, the changes brought by the D component are momentary and due to sudden LC changes only. As a result, envelopes generated in the ID mode are susceptible to having infinite lengths identical to those of the I-mode.

\subsection{The PID Mode}
\label{S33}

\begin{figure}[!b]
\centering
\subfloat[$K_{p} = 10^-3, K_{i}=10^-4, K_{d} = 10^-3$]{\includegraphics[width=0.25\textwidth]{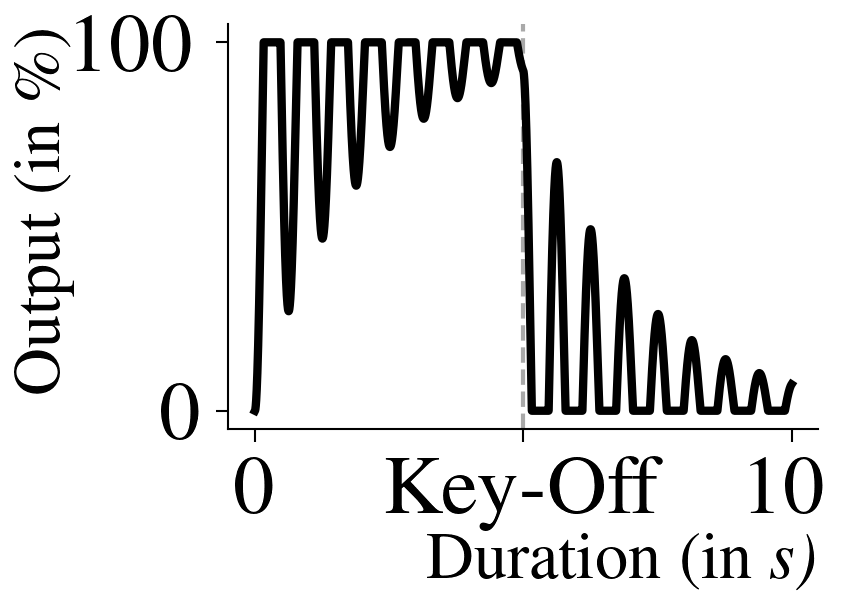}\label{F17a}}
\subfloat[$K_{p} =  10^-2, K_{i}= 10^-4, K_{d} = 10^-3$]{\includegraphics[width=0.25\textwidth]{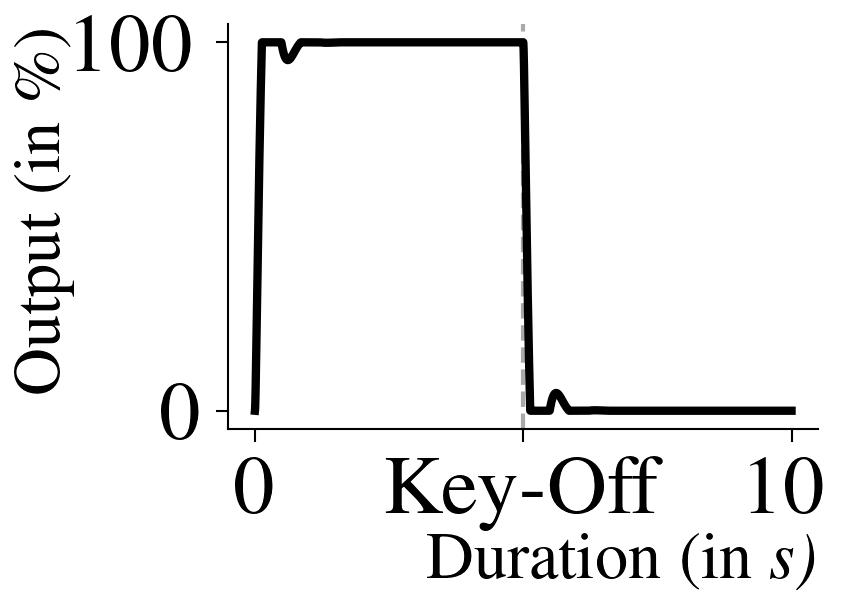}\label{F17b}}\hfill
\subfloat[$K_{p}= 10^-3, K_{i}=10^-3, K_{d} = 10^-3$]{\includegraphics[width=0.25\textwidth]{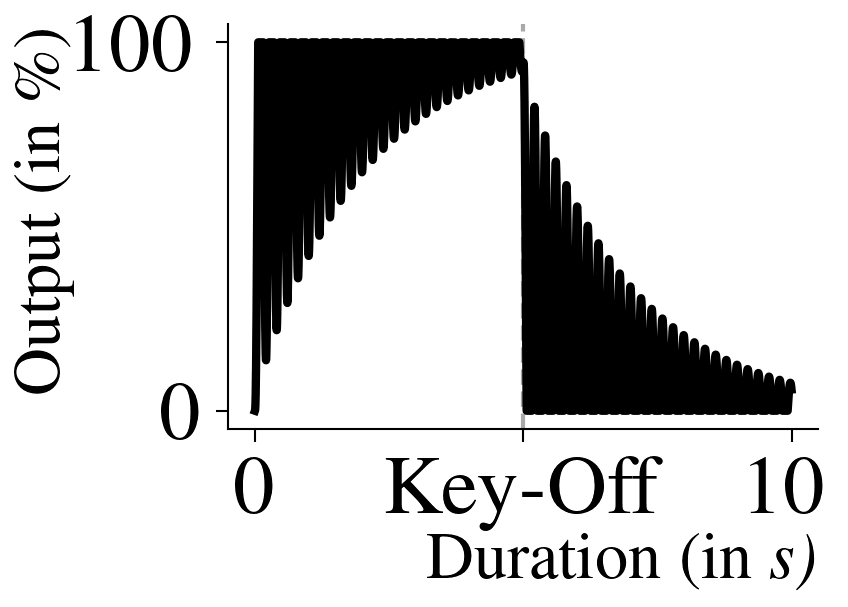}\label{F17c}}
\subfloat[$K_{i}= 10^-3, K_{i}= 10^-4, K_{d}  = 1$]{\includegraphics[width=0.25\textwidth]{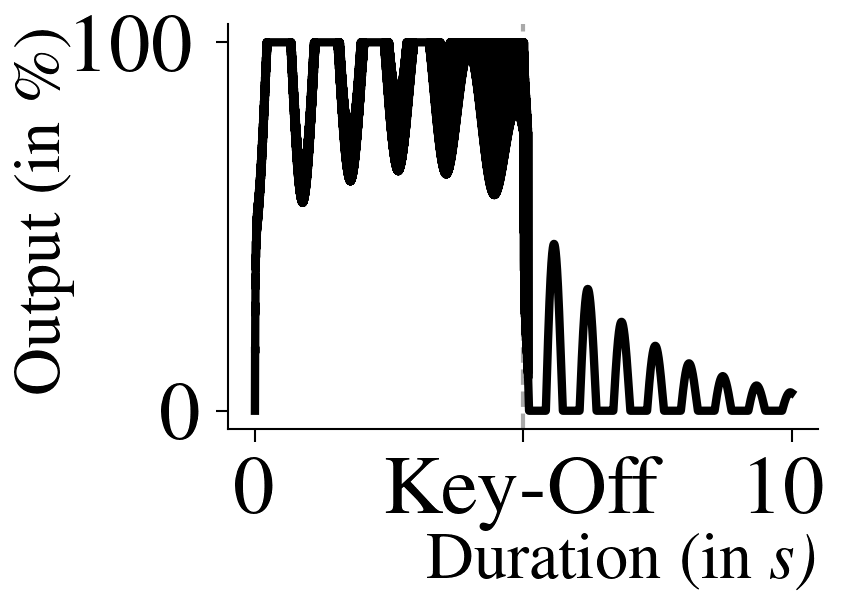}\label{F17d}}\hfill
\caption{Envelopes generated for PIDEGs operating in PID-mode. By combining the effects of $K_{p}$, $K_{i}$ and $K_{d}$ parameters, this mode provides the highest variability in envelope shapes. Since PIDEG operation occurs at full capacity, it is recommended to benchmark them in the PID mode.}
\label{F17}
\end{figure}

As is the case for the dual-controlled modes, the envelopes generated for the triple-controlled or PID-mode are coalescent of the discrete effects of the P, I and D components relative to the value of their respective gains. The contribution of the D component is imperceptible at lower values of $K_{d}$, with the PIDEG appearing to operate in the PI-mode in these situations. As $K_{d}$ increases to higher values, the envelope output reflects sharp deviations on likewise changes in the LC. With an increase in $K_{i}$ values from zero, the oscillations in envelope output have increased frequencies. As is the case for PI- and PD-modes, the $K_{p}$ parameter crafts the basic shape of the envelope about which oscillations occur over time. On increasing it, the envelope saturates quicker at either limit. 

The granularity provided by each control in the PID-mode appreciably increases the possibilities of generating varied envelope shapes, as evidenced by \cref{F17}. PIDEGs in this mode can be considered to be running at full capacity. Therefore, analyzing the performance of the PIDEG algorithm, in terms of the computational complexities involved, and drawing subsequent comparisons with popular ADSR implementations like the STK \cite{scavone} should be done for the PIDEG operating in the PID-mode. However, it is beyond the scope of this research.

\section{Summary}
\label{S4}

ADSR envelopes have greatly influenced the synthesis of music, both in the hardware and software domains. On the other hand, the PID scheme has been a long-term mainstay in feedback-based industrial control applications. The PIDEG framework discussed in this paper attempts to explore the possibilities of using the fundamentals of PID to generate envelopes that have a marked distinction from the standard ADSR— by extrapolating its underlying flexibilities to cause stability and controlled instability conditions preferentially. Through this study, the individual and combinatory effects of the three controlling PIDEG parameters: $K_{p}$, $K_{i}$ and $K_{d}$; as well as the two LC-related parameters, $K_{r}$ and $K_{f}$, were analyzed to conveniently generate envelopes having varying rates of rise and fall, apart from being marked by oscillations that constituted of changing amplitudes and frequencies with time. However, there are certain aspects of PIDEGs where further investigation may be required beyond this study.

\subsection{Future Scope}
\label{S41}

\subsubsection{Replicating ADSR Decay and Sustain}
\label{S411}

In the two-phased LC scheme described in \cref{S21}, the envelope does have the potential to display oscillatory variations on prolonged key-on durations. However, inverse-exponential or trapezoidal LCs may lack the flexibility of dropping and settling the envelope to a desired level (which could be as low as 0 \%) due to their value constancy in this period. This might make it challenging for PIDEGs to model instruments like the kick drum \cite{garcia}, flute \cite{ashtamoorthy} and piano \cite{lee2}. The decay and sustain phases in ADSR, on the other hand, dedicatedly serve to control this nature of the envelope. Therefore, the PIDEG scheme could benefit from adding additional curve phases once the rising inverse-exponential or trapezoidal LC has reached its peak to replicate ADSR decay and sustain modes. Alternatively, using ADSR as an LC source may also suffice, as seen in \cref{F8}.

\subsubsection{Analogue Electronic Implementations}
\label{S412}

The envelopes discussed as part of \cref{S3} were generated using a software implementation of the PIDEG. However, in practice, PID controllers are available in varied forms, including pneumatic and electronic systems \cite{liptak}. While microcontrollers are a suitable option to realize PIDEGs as Eurorack modules or as components of non-modular synthesizers, it would also be interesting to assess the envelopes produced by purely analogue electronic PIDEGs made of capacitors, resistors and op-amps \cite{michal}.

\section{Conclusion}
\label{S5}

In the current form, PIDEGs may be used as an experimental basis to generate audio timbres that could not only resemble those produced by ADSR envelopes but also potentially create novel ones that are beyond its capability. The incorporation of PIDEGs into synthesizer designs could even lead to distinguishing styles of music. Nevertheless, there is enough scope to further examine PIDEG behaviour for the provided parameter inputs, among other areas of possible additions or improvements in the future, before proposing it as a competent alternative to the ADSR envelope generator.

\bibliographystyle{IEEEtran}
\bibliography{citations.bib}

\end{document}